\def\be{\begin{equation}}
\def\ee{\end{equation}}
\documentstyle[eqsecnum,aps]{revtex}
\begin{document}
\hsize=6.5truein\hoffset=0.25truein
\vsize=9.0truein

\draft
\title{New Mesons in the Chirally Symmetric Plasma}
\author{H. Arthur Weldon}
\address{Department of Physics, West Virginia University,
Morgantown, WV 26506-6315}
\maketitle
\begin{abstract}
A nonperturbative proof is given that the chirally-invariant quark propagator contains both particle
and hole singularities with different dispersion relations. Mesons made of
a quark and a hole  will produce dilepton pairs at  masses characteristic of the plasma and with a
distinct energy dependence.
 \end{abstract}
\pacs{12.38.Mh, 11.10.Wx, 25.75.+r}

\section{Introduction}

The central purpose of the experimental program at RHIC is to produce
 a quark-gluon plasma. Despite a great deal of theoretical effort there
is no clear experimental signature of the quark-gluon plasma that will
distinguish it from the very complicated hadron phase. The Theory Workshop
on Relativistic Heavy Ion Collisions is intended both to refine previous ideas
and to initiate discussion of new possibilities. It is in the latter spirit that
this paper proposes   investigation of a new experimental signal not previously
considered.   The present analysis of this signal  still leaves  unanswered questions
that require further investigation by other theorists using other techniques. 

In the QGP at temperature $T$, a typical state  $|\Phi\rangle$ will have
almost all the quark and antiquark modes with momentum less than $T$ occupied
and considerably fewer  modes occupied that have momentum  larger
than $T$. If a quark or  antiquark  is added to  this background the result is
$b^{\dagger}|\Phi\rangle$ or  $d^{\dagger}|\Phi\rangle$. The new possibility is
that removal of a quark gives a state  $b|\Phi\rangle$ and 
the norm of this  quark vacancy state is not small if the momentum is
less than $T$. Similarly an antiquark vacancy is  $d|\Phi\rangle$.
Chiral symmetry plays an essential role in this argument because
the hole states are only important if $b|\Phi\rangle$
has  an energy higher than $|\Phi\rangle$ and this is only possible when the constituent
quark mass is smaller than $T$. Thus approximate chiral symmetry is
essential. This has two consequences: (1) Only $u$ and $d$ quarks are
candidates and (2) When chiral symmetry is badly broken, i.e. the hadron phase,
there are no hole states.  

The proof of hole states in Sec II is entirely non-perturbative. 
The end of Sec II contains references to   perturbative results.
Sec III focusses on experimental consequences of the hole states, the most important of which is
the annihilation of
a quark and hole  into  a virtual photon that subsequently converts to a lepton
pair. If this quark and hole are bound into a meson, there will be a resonance in the dilepton
invariant mass spectrum with new features. 

\section{Quark holes in the exact thermal propagator}

\subsection{Singularities of the exact propagator}

It is remarkable that certain simple properties of the exact thermal quark propagator guarantee that
the existence of low momentum hole states ($k<T$) in the chirally symmetric plasma.
The inverse of the exact time-ordered thermal propagator must be of the form
\be
\bigl[S(\omega,\vec{k})\bigr]^{-1}=\gamma_{0}A(\omega,k)-\vec{\gamma}\cdot\hat{k}B(\omega,k)\ee
because of chiral invariance. The propagator itself is 
\be
S(\omega,\vec{k})={1\over 2}{\gamma_{0}-\vec{\gamma}\cdot\hat{k}\over
A-B}+{1\over 2}{\gamma_{0}+\vec{\gamma}\cdot\hat{k}
\over A+B}.\label{prop1}\ee
At $k=0$ the propagator cannot depend on the direction of $\hat{k}$. Therefore
$B(\omega,0)=0$ and consequently   $S(\omega,0)=\gamma_{0}/A(\omega,0)$. Now comes the one
dynamical input, namely that  at $k=0$ there is a pole in the propagator at the  thermal mass
of the quark. Lattice calculations of temporal quark propagators have measured this thermal mass
\cite{Karsch}.
 The thermal mass must have a negative imaginary part because of damping. By
dimensional analysis, both the real and imaginary parts are proportional to temperature. Denote this
complex mass by $m_{T}$. Then at $k=0$,  $A(\omega,0)=(\omega-m_{T})a(\omega)$ where
$a(\omega)$ does not have a pole at $\omega=m_{T}$. 
At non-zero momentum $A$ has the structure $A(\omega, k)=(\omega-m_{T})a(\omega)+A_{1}(\omega, k)$
where $A_{1}$ is defined to vanish at $k=0$. The first denominator in (\ref{prop1}) is
$A-B=(\omega-m_{T})a(\omega)+A_{1}(\omega, k)-B(\omega,k)$. Since $A_{1}$ and $B$ both vanish at
$k=0$, they are both small when $k$ is chosen sufficiently small. Therefore $A-B$ will vanish at 
$\omega=E_{p}$, where $E_{p}$ is equal to $m_{T}$ plus a small, $k$-dependent correction:
\be
E_{p}\approx m_{T}+\bigl[-A_{1}(m_{T},k)+B(m_{T},k)\bigr]/a(m_{T})+\cdots
\quad (k\ {\rm small})\label{part}\ee
For small $k$ the other denominator  $A+B$ vanishes at $\omega=E_{h}$ where $E_{h}$ is equal to
$m_{T}$ plus a \underbar{different} small $k$ correction because $B$ occurs with opposite sign:
\be
E_{h}\approx m_{T}+\bigl[-A_{1}(m_{T},k)-B(m_{T},k)\bigr]/a(m_{T})+\cdots
\quad (k\ {\rm small})\label{hole}\ee
The first energy will be called the particle energy and the second the hole energy. The
justification for these names will come in the next section. It is easy to see that the inclusion of 
a bare quark mass in the propagator spoils these arguments.

Knowing that  $A-B$ vanishes at $\omega=E_{p}$ and $A+B$ vanishes at
$\omega=E_{h}$, one can  show that  PCT invariance requires $A-B$ to also vanish
at $\omega=-E_{h}$ and $A+B$ to also vanish at $\omega=-E_{p}$. The exact propagator therefore has
four singularities and may be written
 \be
S(\omega,\vec{k})={1\over 2}(\gamma_{0}-\vec{\gamma}\cdot\hat{k})
\Bigl[{Z_{p}(\omega,k)\over \omega-E_{p}}+{Z_{h}(-\omega,k)\over\omega+E_{h}}\Bigr]
+{1\over 2}(\gamma_{0}+\vec{\gamma}\cdot\hat{k})
\Bigl[{Z_{p}(-\omega,k)\over \omega+E_{p}}+{Z_{h}(\omega,k)\over\omega-E_{h}}\Bigr],
\label{prop2}\ee
where Im $E_{p}<0$ and Im $E_{h}<0$. 
It is difficult to deduce much about the numerator functions $Z_{p}$ and $Z_{h}$.
They are very complicated functions of $\omega$ and contain all the branch cuts and any other
singularities of the exact propagator. The $\pm \omega$ in the arguments of these  functions
 are determined by PCT. From the previous argument we know that
at zero momentum the residues of the two kinds of poles are equal:
$Z_{p}(m_{T},0)=Z_{h}(m_{T},0)$.

\subsection{Operator meaning of the hole singularities}

The existence of a pole at a  $m_{T}$ in the zero-momentum propagator  
led to the distinct poles at $\pm E_{p}$ and $\pm E_{h}$ in (\ref{prop1}). The next task is to
figure out what these poles mean in terms of quark creation and destruction operators.
At time t=0 the exact  field operator for massless quarks in a finite volume $V$ can be expanded as 
\be
\psi(0,\vec{x})={1\over\sqrt{V}}\sum_{\vec{k},s}\bigl( u_{\alpha}(\vec{k},s)
b_{\vec{k},s}e^{i\vec{k}\cdot\vec{x}}
+v_{\alpha}(\vec{k},s)d^{\dagger}_{\vec{k},s}e^{-i\vec{k}\cdot\vec{x}}\bigr).\label{psi}\ee
 The field operator at any other time can be expressed as
 $\psi(t,\vec{x})=\exp(iHt)\psi(0,\vec{x})\exp(-iHt)$.
The exact time-ordered thermal propagator may be written in terms of two resolvents:
 \be
S(\omega,\vec{k})={1\over 2}\bigl(\gamma_{0}-\vec{\gamma}\cdot\hat{k}\bigr) 
\Bigl[R_{p}(\omega,k)-R_{h}(-\omega,k)\Bigr]
+{1\over 2}\bigl(\gamma_{0}+\vec{\gamma}\cdot\hat{k}\bigr)
\Bigl(-R_{p}(-\omega,k)+R_{h}(\omega,k)\Bigr)\label{prop3}\ee
\be
R_{p}(\omega,k)\equiv \langle b_{\vec{k}}\;
{1\over \omega+i\epsilon-\hat{H}}\;b^{\dagger}_{\vec{k}}\rangle
\quad
R_{h}(\omega,k)\equiv 
\langle b^{\dagger}_{-\vec{k}}\;
{1\over \omega+i\epsilon -\hat{H}}\;b_{-\vec{k}}\rangle,\label{res}\ee
where PCT was used to express antiquark resolvents in terms of quark resolvents.
For compactness I have used the thermo field dynamics notation
$\hat{H}=H-\tilde{H}$ and the brackets denote the expectation value in the TFD
ground state \cite{Rev}. Both resolvents (\ref{res}) are analytic for  Im $\omega>0$ and have branch
cuts along the real $\omega$ axis.

The pole in (\ref{prop2}) at $\omega=E_{p}$ is in the lower half $\omega$ plane and has a
characteristic spinor structure. Comparison with (\ref{prop3}) shows that it must come from
 $R_{p}(\omega,k)$ since $R_{h}(-\omega)$ is analytic for Im $\omega<0$.
To understand what this means, recall that each state  state $|\Phi\rangle$ in the thermal average is
an energy eigenstate. The existence of the pole at $E_{p}$ means roughly  that the state 
$b^{\dagger}|\Phi\rangle$ has an energy that is higher by $E_{p}$. (More precisely,  
$b^{\dagger}|\Phi\rangle$ is not an energy eigenstate but it has a non-zero overlap with an energy
eigenstate that is higher by $E_{p}$.)

The new feature in (\ref{prop2}) is the pole at $\omega =E_{h}$ in the lower-half $\omega$ plane
with the spinor structure $\gamma_{0}+\vec{\gamma}\cdot\hat{k}$. Comparison with (\ref{prop3})
shows that it can only come from $R_{h}(\omega, k)$ since $R_{p}(-\omega)$ is analytic for Im
$\omega<0$.  Therefore the state $b|\Phi\rangle$ with one fewer particle than $\Phi$
nevertheless has an energy that is higher by $E_{h}$. This state is called a quark hole or quark
vacancy.

The residue of the poles requires evaluating the numerator functions $Z(\omega,k)$ in (\ref{prop2})
at the energy of the pole. At $k=0$ the holes have the same residue as the particles:
$Z_{h}(m_{T},0)=Z_{p}(m_{T},0)$. However at  $k> T$ the residue of the hole
$Z_{h}(E_{h},k)$ becomes very small. This means that high momentum hole states
decouple from the field operator $\psi(x)$. It does not mean that high momentum hole states
disappear, only that $b|\Phi\rangle$ becomes a poor description.  Since gluons couple to color
currents $\overline{\psi}\lambda^{a}\gamma^{\mu}\psi$ and photons couple to  the electromagnetic
current $\overline{\psi}\gamma^{\mu}\psi$, hole states with high momentum propagate  have very
small coupling to gluons and photons. 

\subsection {Perturbative results on holes}

Nothing in the argument presented above is perturbative. 
However the observation  that the thermal quark propagator contains two kinds of poles, one for
quarks and one for holes, did arise in one-loop perturbative calculations  \cite{VV,W,Rob}. The
quark and hole states have been constructed to one-loop order in \cite{BOI}. The dependence of the
spectrum upon chiral symmetry has been studied in  one-loop order \cite{BBS,EP}. The quark and hole
states have been used in calculations of dilepton production  \cite{BPY} and of Higgs damping rates
\cite{Elm}. 

\section{New mesons using hole states}

\subsection{Binding of quarks and holes}

It is easy to envision how new mesons can arise in the chirally symmetric phase though it is
difficult to prove their existence.  If $|\Phi\rangle$ is a color singlet state in the thermal
ensemble then $b^{\dagger}_{k}|\Phi\rangle$ is a color triplet 
with an energy higher by $E_{p}(k)$ after all the interactions between the $b^{\dagger}$ and the
plasma are included.  Similarly $b_{k^{\;\prime}}|\Phi\rangle$ is a color antitriplet with an
energy higher by $E_{h}(k')$ when all its interactions are included. A color singlet state
$b^{\dagger}_{k}b_{k^{\;\prime}}|\Phi\rangle$ should have an energy of approximately
$E_{p}(k)+E_{h}(k')$ minus the color binding energy. 
 
In the chirally symmetric phase the color force is not confining but it is still attractive in
color singlet channels. There are two different contributions to this binding: gluon exchange and
instantons. The  first may be thought of in terms of a Bethe-Salpeter equation
with a kernel for the exchanged gluons. Although the numerators of the quark
propagator (\ref{prop2}) have no mass term, it is the denominators that control the kinematics.
This leads to differential operators that are approximately Klein-Gordon 
$\partial^{2}+m_{T}^{2}$, with $m_{T}\approx$ 150 MeV, acting on the coordinates of the
Bethe-Salpeter wave function. This is  analogous to the positronium problem and even non-relativistic
analysis may be useful. 

A second source of color binding comes from non-trivial topologies, specifically the random
instanton liquid advocated by Shuryak \cite{Shur}. Here because of chiral symmetry the field operator
satifies the equation $\gamma_{\mu}D^{\mu}\psi=0$ with no chiral-symmetry breaking mass term.
 When $A^{\mu}$ is an instanton configuration the classical solutions of this produce a strong
attraction in mesonic channels in which the fermions have the opposite chirality, e.g. attractive in
the scalar channel but not in the vector channel. The lattice calculations of the Shuryak group
confirm this picture. Note that  although the thermal mass of the quarks is present, it is irrelevant
for the instanton attraction. 

\subsection{Quantum numbers of $QQ_{h}$ mesons}

To determine the possible quantum numbers of the new mesons it is useful to examine the bilinear
operators $J_{\Gamma}^{a}(x)=\overline{\psi}_{i}(x)
t^{a}_{ij}\Gamma\psi_{j}(x)$, where $\Gamma$ is one of the sixteen Dirac
matrices and $t^{a}$ are one of the four $2\times 2$ matrices for isospin 0 or 1.
The choice of $\Gamma$ determines the value of $J$ for the operator and its
behavior under parity and charge conjugation. To determine the quark operator
content of each $J_{\Gamma}^{a}$, we again use the  expansion (\ref{psi}) of the exact
field operator. Integrating $J_{\Gamma}^{a}(0,\vec{x})$ over
three dimensional space then gives an operator describing a meson at rest.

\vfill\eject
For example, if $\Gamma=i\gamma_{5}$ then the result is an operator for pseudoscalar
$\pi^{a}$ and $\sigma$ mesons:
\be
\int d^{3}x\overline{\psi}_{i}(\vec{x})t^{a}_{ij}i\gamma_{5}\psi_{j}(\vec{x})
=2i\sum_{\vec{k},s}\bigl[b_{i}^{\dagger}(\vec{k},s)t^{a}_{ij}d_{j}^{\dagger}
(-\vec{k},s)-d_{i}(-\vec{k},s)t^{a}_{ij}b_{j}(\vec{k},s)\bigr].
\ee
This operator has the expected structure: $b^{\dagger}d^{\dagger}$ for the
creation of a quark-antiquark pair plus its PCT conjugate $db$.
The pair has  no orbital angular momentum, L=0, and no spin, S=0. Since
the total angular momentum is J=0, it correctly describes the $\pi$ and $\eta$
mesons as $^{1}S_{0}$ state of quark and antiquark.

As an example of a meson
operator that gives a quark paired with a hole, let $\Gamma=\gamma_{0}$. Then
\be
\int d^{3}x\overline{\psi}_{i}(\vec{x})t^{a}_{ij}\gamma_{0}\psi_{j}(\vec{x})
=2\sum_{\vec{k},s}\bigl[b_{i}^{\dagger}(\vec{k},s)t^{a}_{ij}b_{j}
(\vec{k},s)+d_{i}(\vec{k},s)t^{a}_{ij}d_{j}^{\dagger}(\vec{k},s)\bigr].
\ee
If this  operator acts on a typical state $|\Phi\rangle$ containing many quarks
and antiquarks it can create a new kind of meson at rest,   
$b^{\dagger}(\vec{k})b(\vec{k})|\Phi\rangle$ (or the PCT conjugate) with an energy approximately 
$E_{p}+E_{h}$ minus the binding energy. 

The chart below tabulates the quantum numbers of the possible mesons 
for all 16 Dirac matrices. The choice of $\Gamma$ determines the value of $J^{PC}$ for that
operator. For the last four operators in the chart, $J_{\Gamma}^{a}$ contains 
both ordinary meson combinations $b^{\dagger}d^{\dagger}$ and new combinations $b^{\dagger}b$.

\be\matrix{\Gamma 
&J^{PC} 
&\qquad Q\overline{Q} \ {\rm MESONS}
&\qquad QQ_{h} \ {\rm MESONS}\cr\cr
i\gamma_{5}
& 0^{-+}
&\qquad ^{1}S_{0}: \pi\ {\rm and}\ \eta 
&\qquad --\cr
&&\hskip4em\swarrow\searrow&\cr
1&0^{++}
&\qquad ^{3}P_{0}: a_{0}\ {\rm and}\ \sigma 
&\qquad --\cr\cr 
\gamma_{0}
&0^{+-}
&\qquad --
&\qquad ^{1}S_{0}: {\rm \ I=1,\; 0}\cr
&&&\hskip5em\downarrow\cr 
\gamma_{5}\gamma_{0} 
& 0^{-+}
&\qquad -- 
&\qquad ^{3}P_{0}: {\rm \ I=1, \; 0}\cr\cr
 \vec{\gamma}
& 1^{--} 
&\qquad ^{3}S_{1}: \rho\  {\rm and} \ \omega 
&\qquad ^{1}P_{1}:  {\rm I=1, \; 0}\cr
&&\hskip1em\downarrow
&\hskip5em\downarrow\cr 
\gamma_{5}\vec{\gamma}
& 1^{++} 
&\qquad ^{3}P_{1}: a_{1}\  {\rm and} \ f_{1}
&\qquad ^{3}D_{1}: {\rm I=1, \; 0}\cr\cr
\gamma_{5}\gamma_{0}\vec{\gamma}
&1^{+-} 
&\qquad ^{1}P_{1}: b_{1}\  {\rm and} \ h_{1}
&\qquad  ^{3}S_{1}: {\rm I=1, \; 0}\cr
&&\hskip4em\swarrow\searrow
&\hskip6em\swarrow\searrow\cr
\gamma_{0}\vec{\gamma}
& 1^{--} 
&\qquad ^{3}D_{1}: \rho^{\,\prime}\ {\rm and}\ \omega'
&\qquad ^{3}P_{1}: {\rm I=1, \; 0}\cr}\ee

\noindent 
The lightest $Q\overline{Q}$ mesons are at the top of the chart; the heaviest at the
bottom. Isotriplets are listed before isosinglets. 
The arrows in the chart indicate the opposite parity states that are degenerate in mass because of
SU(2)$_{A}$ symmetry. The SU(2)$_{A}$ multiplets have various dimensions:  $\pi+\sigma$ 
is four dimensional;  $\rho+ a_{1}$  is six dimensional; $\omega$ is a chiral singlet.
The $Q\overline{Q}$ mesons automatically have $P=(-1)^{L+1}$ and $C=(-1)^{L+S}$. 
The $QQ_{h}$ mesons in the second column have $P=(-1)^{L}$ but $C$ is not determined by the values
of $L$ and $S$.

\subsection{Decays and experimental signals of $QQ_{h}$ mesons}

In the chirally invariant phase not all the $Q\overline{Q}$ and  $QQ_{h}$ mesons will be bound.
The combinations that are bound into isosinglets have a Zweig suppressed  decay into two or more
gluons depending upon  $J^{PC}$. Their decay into photons might be detectable. The isotriplets cannot
decay into gluons but they could break up by collisions with free quarks in the plasma. Most
importantly, the holes
 can only exist in the chirally invariant
phase. Thus in a RHIC collision they can  exist for a maximum of 5 or 10 fm/c, corresponding to a
minimum width of  20 MeV to 40 MeV. 

In the random instanton liquid  model \cite{Shur} the combinations that bind  are those whose
constituents have opposite chirality, viz $\Gamma=1, i\gamma_{5}, 
\gamma_{5}\gamma_{0}\vec{\gamma}, \gamma_{0}\vec{\gamma}$. The last two of these would give $QQ_{h}$
vector mesons with $J^{PC}=1^{+-}$ and $1^{--}$. The latter  would produce  a characteristic
dilepton pair signal at its mass. It seems quite unlikely that this thermal meson mass would
coincide with that of any known sources of dileptons. 

The calculation of meson binding and of decay depends crucially on  the residue
function $Z_{h}(E_{h},k)$. As mentioned earlier  it is the same as the quark residue $k=0$ but at 
$k>T$ it is negligible. This complication helps distinguish the $QQ_{h}$ dilepton signal
from background. All dilepton rates 
fall exponentially like $\exp(-E/T)$ where $E$ is the energy of the lepton pair. 
For normal sources such as $\rho$ and $\omega$, the prefactor of this exponential is
almost independent of $E$.  For dileptons produced by $QQ_{h}$ mesons the prefactor will
 fall rapidly with the dilepton energy $E$ and be negligible at $E\sim T$.
Thus in a plot of dilepton invariant mass that includes only low energy pairs
 such as $E<$ 50 MeV or $E<100$ MeV, the $QQ_{h}$ signal is most likely to stand out.   

\acknowledgments

It is pleasure to thank Sid Kahana and the entire BNL theory group for organizing and hosting 
 a very stimulating workshop.
This work was supported in part by National Science Foundation grants
PHY-9213734 and PHY-9630149.

ed. R.C. Hwa (World Scientific, Singapore).

\bibitem{BBS} G. Baym, J.P. Blaizot, and B. Svetitsky, Phys. Rev. {\bf D 46}, 4043 (1992).

\bibitem{EP} E. Petitgirard, Z. Phys. {\bf C 54}, 673 (1992).

\bibitem{BPY} E. Braaten, R.D. Pisarski, and T.C.Yuan, Phys. Rev. Lett. {\bf 64} 2242 (1990).

\bibitem{Elm} P. Elmfors, K. Enqvist, and I. Vilja, Nucl Phys. {\bf B 412}, 459 (1994).

\bibitem{Shur} E. V Shuryak and J.J.M. Verbaarschot, Nucl. Phys. {\bf B 410}, 37 and 55 (1993);
T. Sch\"{a}fer, E. V Shuryak, and  J.J.M. Verbaarschot, Nucl. Phys. {\bf B 412}, 143 (1994).

\end{document}